\documentclass{article}

\usepackage[preprint]{neurips_2025}

\usepackage[utf8]{inputenc} 
\usepackage[T1]{fontenc}    
\usepackage{hyperref}       
\usepackage{url}            
\usepackage{booktabs}       
\usepackage{amsfonts}       
\usepackage{nicefrac}       
\usepackage{microtype}      
\usepackage{xcolor}         
\usepackage{amsmath}
\usepackage{array}
\usepackage{graphicx }
\usepackage[notransparent]{svg}
\usepackage{booktabs}
\usepackage{multirow}
\usepackage{subcaption} 
\usepackage{url}
\usepackage{lipsum}
\usepackage{wrapfig}
\usepackage{algorithm}
\usepackage{algpseudocode}    

\usepackage{tikz}
\usetikzlibrary{arrows.meta,arrows}
\title{
Ken Utilization Layer: Hebbian Replay Within a Student’s Ken for Adaptive Exercise Recommendation

}

\author{
  Grey Kuling \\
  Department of Biomedical Informatics \\
  Curriculum Fellows Program \\
  Harvard Medical School \\
  \texttt{grey\_kuling@hms.harvard.edu} \\
  \And
  Marinka Zitnik \\
  Department of Biomedical Informatics \\
  Harvard Medical School \\
  Kempner Institute, Broad Institute\\
  \texttt{marinka@hms.harvard.edu} \\
}

\begin{document}
\maketitle

\begin{abstract}
Adaptive exercise recommendation (ER) aims to choose the next activity that matches a learner’s evolving Zone of Proximal Development (ZPD). We present KUL-Rec, a biologically inspired ER system that couples a fast Hebbian memory with slow replay-based consolidation to enable continual, few-shot personalization from sparse interactions. The model operates in an embedding space, allowing a single architecture to handle both tabular knowledge-tracing logs and open-ended short-answer text. We align evaluation with tutoring needs using bidirectional ranking and rank-sensitive metrics (nDCG, Recall@K). Across ten public datasets, KUL-Rec improves macro nDCG (0.316 vs. 0.265 for the strongest baseline) and Recall@10 (0.305 vs. 0.211), while achieving low inference latency and an $\approx99$\% reduction in peak GPU memory relative to a competitive graph-based model. In a 13-week graduate course, KUL-Rec personalized weekly short-answer quizzes generated by a retrieval-augmented pipeline and the personalized quizzes were associated with lower perceived difficulty and higher helpfulness (p < .05). An embedding robustness audit highlights that encoder choice affects semantic alignment, motivating routine audits when deploying open-response assessment. Together, these results indicate that Hebbian replay with bounded consolidation offers a practical path to real-time, interpretable ER that scales across data modalities and classroom settings.

\end{abstract}

\section{Introduction} 

Recommending the right exercise at the right time is a cornerstone of effective personalized education. The goal is to dynamically select questions or activities that match a student’s current ability, reinforce prior knowledge, and introduce appropriate challenge. This task, exercise recommendation (ER), requires fine-grained modeling of what a learner knows and how they learn, ideally based on only a few examples. Traditional systems address this by first inferring a learner's knowledge state, typically using knowledge tracing (KT) models, and then selecting content based on estimated mastery \citep{piech2015deep,zhang2017dynamic}. While this two-step pipeline has driven advances in intelligent tutoring systems, it hinges on a crucial assumption: that student knowledge can be reliably inferred from sparse data using cohort-trained models.

However, modern KT algorithms often require extensive training on large, labeled student datasets to achieve robust performance \citep{sonkar2020qdkt,liu2023xes3g5m}. These models struggle to personalize effectively when evidence is limited, especially for students with atypical learning patterns or those entering from outside the training distribution \citep{wang2020generalizing,bai2025cskt}. Worse, they tend to collapse under curricular drift, where skills evolve over time or differ between institutions. This makes them ill-suited for exercise recommendation in real-world educational settings, where systems must adapt on-the-fly to sparse, individualized data, often without the luxury of large cohort histories.

Three challenges are central to this problem. First, KT models trained on historical data often fail to adapt quickly or accurately to new students, limiting their value in cold-start scenarios. Second, forgetting is rarely modeled in a cognitively meaningful way; instead, decay is managed through ad hoc heuristics or replay buffers, adding computational overhead \citep{kurth2023replay,jeeveswaran2023birt,batra2024evcl}. Third, most KT systems rely on structured input formats with predefined topic labels (e.g., “fractions,” “grammar rules”), limiting their ability to handle richer, more authentic data types such as short-answer text or open-ended reasoning \citep{abdelrahman2023knowledge,zhou2024predictive}.

We reframe the problem from predicting student correctness to selecting appropriate exercises for individual learners in real time, and we ground the solution in Complementary Learning Systems (CLS) theory. CLS posits a dual-memory architecture consisting of a fast-learning hippocampal system and a slower-learning neocortical system, balancing rapid adaptation with long-term generalization \citep{mcclelland1995there,kumaran2016learning}. Prior attempts to operationalize CLS in educational AI have often relied on complex architectures and backpropagation-based training, which limits scalability and interpretability \citep{ke2021achieving,wang2024comprehensive}. We propose KUL-Rec, built around a Ken Utilization Layer that couples learned linear transformations with a modern Hopfield associative memory. The memory is updated via a time-variant Hebbian rule with decay, a local and biologically plausible alternative to backpropagation \citep{centorrino2024modeling,remme2021hebbian}. To address sparse supervision in education settings, we introduce a Loss-aligned Internal Target method that analytically computes an ideal internal representation for each desired output, enabling continual personalization from limited feedback such as correctness signals.

KUL-Rec supports both tabular logs and embedded short-answer responses, enabling exercise recommendation from minimal evidence and without reliance on cohort-wide training. Unlike traditional KT models, KUL-Rec does not store raw data, avoids replay, and adapts continually as students interact with the system. This makes it especially suited for practical deployment in settings where privacy, compute, or data availability is constrained.

Empirically, KUL-Rec outperforms deep KT baselines on ten public datasets when re-evaluated for exercise ranking, and shows strong generalization to graduate-level quizzes with open-text responses. Our contributions are fourfold:
(i) a biologically inspired exercise recommender with Hebbian-updated Hopfield memory supporting continual personalization;
(ii) a new analytical method for generating internal training targets from sparse supervision;
(iii) demonstrated cross-modal generalization to diverse educational inputs with reduced computational cost and data requirements; and
(iv) an ER-aligned evaluation protocol that extends standard rank metrics with a bidirectional ranking criterion and a percentile-based ZPD selection rule, plus reproducible candidate-bank construction for KT logs.

\section{Related Work}
\label{sec:background}

\textbf{Exercise recommendation and knowledge tracing signals.\;}
ER aims to rank and select the next learning activity that best aligns with a learner’s evolving knowledge state. Many systems derive internal estimates of learner understanding from KT models, which track mastery over time based on past interactions \citep{piech2015deep}. Classical Bayesian KT framed knowledge acquisition as a probabilistic transition from \emph{unlearned} to \emph{learned} \citep{corbett1994knowledge}, later extended with temporal and contextual factors from item response theory and performance factor analysis \citep{baker2008more,khajah2014integrating}. Deep learning approaches—ranging from recurrent to memory-augmented, transformer-based, and state-space KT—leverage large-scale interaction logs to uncover complex temporal patterns in learning behavior \citep{sonkar2020qdkt,zhang2017dynamic,long2021tracing,ghosh2020context,cao2024mamba4kt,zhou2025revisiting,liu2025deep,yang2023evolutionary}.

However, these models are typically optimized to predict student correctness at a cohort level, not to rank content for individuals. Large-scale KT benchmarks such as XES3G5M \citep{liu2023xes3g5m} reflect aggregate performance gains, but do not capture the nuanced demands of ER, which requires fine-grained, real-time adaptation at the individual level. In this work, we repurpose KT signals as tools for ER, prioritizing rank-sensitive recommendation over cohort-level prediction accuracy.

\textbf{Few-shot personalization and the cold-start problem.\;}
A key limitation of cohort-trained KT models is their inability to adapt rapidly and accurately to new learners with limited historical data. This ``cold-start’’ problem is especially acute in personalized learning environments, where systems must operate under sparse, streaming conditions. Several strategies have attempted to improve few-shot adaptation—e.g., constraint-based KT (csKT) \citep{bai2025cskt} and simplification-based methods like SimpleKT \citep{liu2023simplekt}—but these still treat learning as transitions between discrete knowledge concept (KC) identifiers. This abstraction hinders rapid personalization, as it struggles to accommodate out-of-distribution content or nuanced learner behaviors. For ER, the central challenge is updating a learner’s internal representation from minimal evidence, without expensive re-training or storing large volumes of interaction data.

\textbf{Biologically inspired memory and Hebbian learning.\;}
Complementary Learning Systems (CLS) theory posits that the brain uses two interacting memory systems: a fast-learning hippocampus for episodic memory and a slow-learning neocortex for long-term generalization \citep{mcclelland1995there,kumaran2016learning}. This dual-system framework has inspired AI models that combine fast episodic storage with slower consolidation mechanisms, offering a compelling solution for online adaptation in ER. Modern Hopfield networks extend this concept by enabling content-based recall in high-dimensional space with associative memory updates \citep{remme2021hebbian,centorrino2024modeling,molter2005introduction,fachechi2019dreaming}. Recent work revisits CLS principles at scale—e.g., bio-inspired replay to mitigate forgetting \citep{jeeveswaran2023birt}, transformer interpretations of memory consolidation \citep{kim2023transformer}, and continual learning through regularization \citep{batra2024evcl,bian2024make,qi2024interactive}.

In KT and ER contexts, some dual-memory architectures simulate CLS by combining episodic memory banks with periodic distillation into neural representations \citep{ke2021achieving,wang2024comprehensive}. However, these systems often depend on full-data replay or global gradient control, increasing computational burden and deviating from the sparse, local learning mechanisms envisioned by CLS. Hebbian learning offers a biologically grounded alternative: it updates connections based on co-activation patterns and supports continual adaptation under streaming constraints, making it attractive for ER in educational settings.

\textbf{Beyond KC identifiers: richer input modalities and unified representations.\;}
Traditional KT datasets encode interactions as triples of the form \texttt{(student, KC, correctness)}, abstracting away richer data that could improve learner modeling and recommendation. Emerging approaches explore graph-based KT to incorporate prerequisite structures and domain hierarchies (e.g., GKT \citep{nakagawa2019graph}, Graph Transformers \citep{lee2024transitivity,canturk2023graph}), or learn from domain-aware structures \citep{zhou2024predictive}. Still, modalities such as short-answer responses, diagrams, and verbal explanations remain underutilized, despite evidence that they contain valuable signals about student understanding and misconceptions \citep{li2024explainable}. For ER systems to adapt effectively, it is crucial to handle both structured and unstructured data, ideally through a unified embedding space. Embedding-based approaches offer a path forward, supporting both classical KC-based learning and richer open-ended responses in a common framework \citep{carmon2023automated}.

\textbf{Evaluation protocol aligned with ER objectives.\;}
Building on prior work that evaluates educational recommenders with rank-aware metrics \citep{yu2021mooccubex}, we adopt nDCG and Recall@K to assess next-item quality for individual learners. We extend this practice with a bidirectional ranking criterion that conditions the sort order on the observed outcome (descending if correct, ascending if incorrect), rewarding appropriate confidence for items at or beyond the learner’s current reach. For deployment, we pair this with a percentile-based ZPD selection rule. Together, these choices align offline evaluation on KT logs and online selection with tutoring goals under sparse, evolving supervision.

\begin{figure}[ht]
    \centering 
    \includegraphics[width=\linewidth]{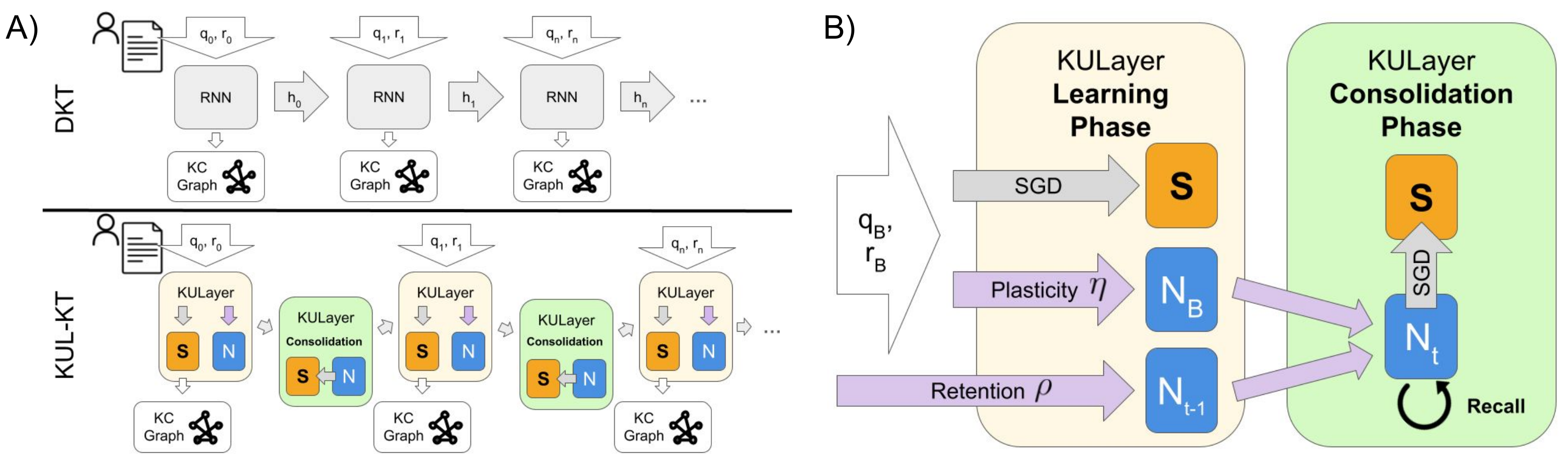} 
    \caption{\textbf{Overview of KUL-Rec Modeling: A)} A learner's longitudinal interactions with quiz questions generate sequential data typically analyzed using recurrent neural networks (RNNs) to output KC graphs of the likelihood the student will get KCs correct given the current learning state. Our proposed model augments this framework by incorporating biologically inspired dual-process memory mechanisms, enabling continual adaptation, rapid instance learning, and generalized knowledge consolidation. \textbf{B) }Detailed view of the Learning and Consolidation phases. During Learning, the model rapidly stores interactions using a Hebbian update with weight decay. Each mini-batch \( B \) consists of questions \( q_B \) and responses \( r_B \), which are encoded into the Hebbian memory module \( N \), forming an update \( N_B \) at time step \( t \). During Consolidation, stored memories are recalled and used to refine representations in the Student network \( S \).
} \label{fig:overview} 
\end{figure}

\section{Materials and Methods}
\label{sec:mandm}

\subsection{Problem Statement}

ER refers to the task of selecting the next learning activity whose difficulty and content are well matched to a learner’s current level of understanding \cite{shabani2010vygotsky}.  
An effective ER system should avoid recommending items that are either too easy or too difficult, and instead identify exercises situated within the learner’s ZPD, where challenge fosters growth without inducing frustration.

At each interaction step \( t \), the learner’s history is recorded as
\[
\mathcal{H}_t = \{(q_\tau, r_\tau)\}_{\tau=1}^{t},
\]
where \( q_\tau \) is an exercise and \( r_\tau \in \{0,1\} \) indicates correctness.  
Given a candidate set of future exercises \( \mathcal{C}_{t+1} = \{ q_{t+1}^{(1)}, \ldots, q_{t+1}^{(K)} \} \), including the item actually shown to the learner, the ER model estimates the probability of success for each:
\[
\hat{r}_{t+1}^{(k)} = P(r_{t+1}^{(k)} = 1 \mid q_{t+1}^{(k)}, \mathcal{H}_t), \quad k = 1, \ldots, K.
\]

These probabilities reflect the model’s belief about which items fall within the learner’s reach. The pedagogical goal is not to maximize correctness, but to recommend exercises near the threshold of mastery—those that are challenging yet attainable.

In some educational settings, exercises are organized according to prerequisite structures or conceptual dependencies. Prior work has modeled these using static graphs, where nodes represent questions and edges represent fixed transitions \citep{nakagawa2019graph,yang2021gikt}.  
To enable continual, learner-specific adaptation, we extend this view with a dynamic transition function:
\[
\hat{e}_{i,j}(t+1) = P(\text{correct}(q_j) \mid q_i, r_i, s_{t+1}) = P(\text{correct}(q_j) \mid \mathcal{H}_{t+1}),
\]
where \( s_{t+1} \) is the updated learner state after observing interaction \( (q_i, r_i) \). This allows the model to reweight conceptual relationships in real time, rather than relying on fixed curricular graphs.

While not central to our evaluation, this dynamic formulation supports future extensions to curriculum planning, path optimization, or sequence-aware recommendation strategies.

\subsection{KUL-Rec: A Memory-Augmented Architecture for Personalized Exercise Recommendation}

To support real-time, individualized recommendation with limited supervision, we introduce KUL-Rec, a neural architecture designed for continual learning from sparse learner interactions. The model is motivated by cognitive theories of human memory and combines a slow-learning linear network with a fast-updating associative memory. KUL-Rec does not require raw data storage or cohort-level training and is suitable for adaptive educational settings where feedback is limited and data arrives incrementally.

Each learner interaction, such as a quiz attempt or written response, is first embedded into a vector representation \( \mathbf{x} \in \mathbb{R}^{d} \). This input passes through a sequence of learned transformations, consisting of two nonlinear layers and one linear output layer. The transformations are defined as follows:
\[
\mathbf{z}_1 = \mathbf{W}_{in} \mathbf{x}, \quad 
\mathbf{h}_1 = \phi(\mathbf{z}_1), \quad 
\mathbf{z}_2 = \mathbf{S} \mathbf{h}_1, \quad 
\mathbf{h}_2 = \phi(\mathbf{z}_2), \quad 
\mathbf{z}_3 = \mathbf{W}_{out} \mathbf{h}_2
\]
where \( \phi(\cdot) \) is the GELU activation function. The intermediate representation \( \mathbf{z}_2 \) reflects the learner’s current internal state and interfaces with a memory module responsible for storing and replaying learning traces.

\subsection{Hebbian Associative Memory and Continual Replay}

The associative memory module in KUL-Rec is modeled after the biologically inspired Go-CLS framework, which encodes new experiences via Hebbian learning \citep{sun2023organizing}. In our adaptation, we use this mechanism to support continual, streaming updates of learner interactions without storing raw history or relying on fixed memory slots. The memory system, referred to as the Notebook, enables rapid encoding of transient learning traces and plays a central role in facilitating student-specific generalization over time.

At each step, a sparse binary indexing matrix \( \boldsymbol{\Xi} \in \{0,1\}^{M \times B} \) is generated to represent the currently active subset of memory units, where \( B \) is the mini-batch size and \( M = 2048 \) is the total number of memory units. Each column \( \boldsymbol{\xi}^\mu \) has fixed activity level \( a = 0.05 \), and there are no persistent slot assignments across batches. The input activations \( \mathbf{Z}_1 \in \mathbb{R}^{d_1 \times B} \) and target internal representations \( \mathbf{Z}_2^* \in \mathbb{R}^{d_2 \times B} \) are projected into memory via Hebbian outer products, generating a per-batch memory update \( \mathbf{N}_{\mathcal{B}} \) composed of five matrices:
\begin{equation*}
\begin{aligned}
\mathbf{U}_{z_1 \to \text{N}} &= (\boldsymbol{\Xi} - a)\mathbf{Z}_1^\top  & \quad
\mathbf{U}_{z_2^* \to \text{N}} &= (\boldsymbol{\Xi} - a)\mathbf{Z}_2^*{}^\top \\
\mathbf{V}_{\text{N} \to z_1} &= \frac{\mathbf{Z}_1 (\boldsymbol{\Xi}^\top - a)}{M a(1 - a)} & \quad
\mathbf{V}_{\text{N} \to z_2^*} &= \frac{\mathbf{Z}_2^* (\boldsymbol{\Xi}^\top - a)}{M a(1 - a)} \\
  \multicolumn{4}{c}{%
    $\displaystyle
      \mathbf{J}_{ij} =
      \begin{cases}
        \bigl(\frac{(\boldsymbol{\Xi}-a)(\boldsymbol{\Xi}-a)^\top}{M a(1 - a)}-\frac{\gamma}{M a}\bigr)_{ij}, 
          & i\neq j,\\
        0, & i = j,
      \end{cases}$
  }
\end{aligned}
\end{equation*}
Here, \( \gamma = 0 \) disables lateral inhibition across memory units, and the thresholds are automatically computed following the method of Sun et al \cite{sun2023organizing}.

The global memory matrix \( \boldsymbol{N}_t \) is updated via exponential decay:
\[
\boldsymbol{N}_t = \rho \boldsymbol{N}_{t-1} + \eta \mathbf{N}_{\mathcal{B}},
\]
where \( \rho \in (0,1) \) is the decay rate controlling forgetting and \( \eta \in (0,1) \) is the learning rate controlling plasticity. This update rule ensures that memory evolves with the learner, gradually discarding obsolete traces while reinforcing recently observed associations.

At scheduled intervals (e.g., epoch boundaries), the model enters a consolidation phase. In this phase, sparse vectors \( \mathbf{h}^{(0)} \) are initialized and evolved through recurrent updates using the lateral connectivity matrix \( \mathbf{J} \). The memory trace is recalled using the following update rule:
\[
\mathbf{h}^{(u)} = f(\mathbf{J} \mathbf{h}^{(u-1)} - \theta),
\]
where \( f(\cdot) \) is a nonlinear activation function such as thresholded ReLU and \( \theta \) controls sparsity. The retrieved traces \( \mathbf{h}^{(U)} \) are projected back into the student network space via the readout matrices \( \mathbf{V}_{\text{N} \to z_1} \) and \( \mathbf{V}_{\text{N} \to z_2^*} \). These replayed activations serve as targets for a mean squared error loss used to update the slower-learning student weights. This process mimics biological memory replay and enables the student module to internalize structural regularities observed in recent interactions, without requiring full gradient backpropagation through time.

\subsection{Loss-Aligned Internal Target for Memory Encoding}

Because the memory system operates on internal network representations that are not directly supervised, we must construct surrogate targets that reflect the direction of the learning signal. To this end, we introduce the Loss-aligned Internal Target (LIT), a mechanism that computes ideal internal representations using downstream gradients without storing ground truth output vectors. The LIT method allows memory updates to be both biologically plausible and mathematically aligned with error reduction.

To derive the LIT, we begin with a simplified linear network composed of two layers:
\[
z_1 = W_1 x, \quad \hat{y} = W_2 z_1,
\]
where \( \hat{y} \) is the predicted output and the loss is measured using mean squared error. The goal is to find an adjusted internal representation \( z_1^* \) such that the updated prediction \( \hat{y}^* = W_2 z_1^* \) equals the true label \( y \), eliminating the prediction error. If we assume a linear correction of the form
\[
z_1^* = z_1 - G \nabla_{z_1} \mathcal{L},
\]
then solving for the gain matrix \( G \) leads to the exact form
\[
G = \frac{N}{2}(W_2 W_2^\top)^{-1},
\]
where \( N \) is the batch size. This solution ensures that the corrected representation projects to the desired output in a single update step.

However, the matrix \( W_2 W_2^\top \) may be ill-conditioned or non-invertible in practical settings. To address this, we approximate its inverse using a truncated Neumann series and introduce a stability constant \( \alpha \):
\[
G \approx \frac{N}{2\alpha} \left( I - \frac{1}{\alpha} W_2 W_2^\top \right),
\]
where \( \alpha > \|W_2 W_2^\top\|_{\max} \). This provides a numerically stable and computationally efficient method to compute an ideal correction direction.

This derivation generalizes to deeper networks. In KUL-Rec, the relevant transformation is the output projection matrix \( \mathbf{W}_{out} \), and the surrogate internal representation at the memory interface is given by
\[
\mathbf{z}_2^* = \mathbf{z}_2 - \frac{N}{2\alpha}
\left(I - \frac{1}{\alpha} \mathbf{W}_{out} \mathbf{W}_{out}^\top \right) \nabla_{\mathbf{z}_2} \mathcal{L}.
\]
This target is used in Hebbian memory updates to ensure that the stored representations reflect productive learning directions. The use of LIT enables learning from sparse supervision by approximating the ideal memory content that would lead to reduced error, without requiring full backpropagation or explicit labels at intermediate layers.

Together, the Hebbian memory module and the LIT mechanism allow KUL-Rec to continually adapt to new learners and tasks using biologically motivated learning principles, while maintaining alignment with modern error-minimization objectives.

\subsection{Training and adaptation procedure}
\label{sec:train_adapt}

We adopt a rolling mini-quiz protocol that mirrors classroom pacing. Each learner’s timeline is divided into non-overlapping windows of length \(W=10\) interactions. At step \(k\), the model is updated on window \(k\) using the task loss and immediately evaluated on window \(k{+}1\), so parameters evolve continually and are never reset. After each window update, the associative memory writes the current internal pair \((\mathbf{h}_1, \mathbf{z}_2^{\ast})\) using the Hebbian rule with decay parameters \((\rho, \eta)\). At the end of a window, a short consolidation phase recalls sparse traces by iterating the lateral matrix \(\mathbf{J}\) over a small number of steps, projects recalls back to the student space, and fine-tunes the slow learner with a reconstruction loss. This procedure realizes a complementary learning systems process of rapid episodic capture, replay-based consolidation, and graceful forgetting under streaming constraints.

\subsection{Input Representations and Modalities}

KUL-Rec is compatible with both structured and unstructured educational data formats. For datasets based on traditional tabular logs, each learner interaction is encoded using a discrete embedding that combines question identity and correctness. Each input \( q_t \) and outcome \( r_t \in \{0,1\} \) is mapped to a unique index and transformed into a learned vector representation.

For open-ended inputs such as short-answer responses, we use fixed-length text embeddings obtained from a pretrained language model, specifically OpenAI’s \texttt{text-embedding-ada-002}. These embeddings provide vector representations for both questions and responses, enabling KUL-Rec to process more complex learner input beyond correctness labels. In this setting, the model produces a predicted response vector that can be compared to known answer embeddings using cosine similarity or other distance metrics.

This flexibility allows KUL-Rec to serve as a unified recommendation system across a variety of input formats, supporting both binary outcome prediction and semantic response modeling within the same architectural framework.

\section{Experiments}
\label{sec:experiments}

We evaluate KUL-Rec on two experimental settings. In Section~\ref{sec:tabulardata}, we benchmark the model on ten public tabular knowledge tracing datasets, comparing its performance against sequential, memory-augmented, transformer, and graph-based models using rank-sensitive metrics. In Section~\ref{sec:shortansdata}, we test the model's generalization in a live graduate-level classroom, where KUL-Rec personalizes open-ended short-answer quizzes for individual students. These experiments assess the model for few-shot personalization, continual learning, and adaptive content selection on structured and unstructured educational data.

\subsection{Tabular Data Experiment}
\label{sec:tabulardata}

We benchmark KUL-Rec on ten publicly available educational datasets, each consisting of longitudinal $(q_t, r_t)$ interaction sequences across a diverse set of learners, topics, and item pools. While originally curated for correctness prediction tasks, these datasets are well suited to ER for three reasons. First, they record the actual next item attempted by the learner at each step, enabling an offline evaluation protocol where the observed item serves as the held-out target within a candidate set. Second, they include long per-learner sequences, allowing us to assess continual adaptation, memory retention, and forgetting over time. Third, these datasets have been widely used in prior research, with publicly available splits and baselines that support reproducibility \citep{piech2015deep, liu2023xes3g5m}.

\subsubsection{Datasets and Preprocessing}

The ten datasets span various educational levels and domains, including algebra, general mathematics, and science education. Summary statistics are shown in Table~\ref{tab:dataset_stats1}, including the number of students, interactions, and distinct concepts (or question types) in each dataset.

\begin{table}[h]
\centering
\caption{Summary statistics for benchmark datasets used in the tabular experiment.}
\label{tab:dataset_stats1}
\begin{tabular}{lrrr}
\toprule
\textbf{Dataset} & \textbf{\# Students} & \textbf{\# Interactions} & \textbf{\# Concepts} \\
\midrule
Algebra2005    & 574   & 0.8M    & 112  \\
Bridge2006     & 1146   & 3.6M    & 493     \\
ASSIST2009     & 4217   & 0.3M   & 123    \\
ASSIST2012     & 29018   & 4.6M    & 265    \\
ASSIST2015     & 19840 & 2.1M    & 100    \\
ASSIST2017     & 1709  & 1.0M    & 102    \\
EdNet (small)  & 5000 & 0.9M   & 141    \\
EdNet (large)  & 50000 & 9.8M    & 141    \\
NIPS34         & 4918    & 1.6M   & 62     \\
XES3G5M        & 14453 & 5.8M   & 865   \\
\bottomrule
\end{tabular}
\end{table}

We applied a shared preprocessing pipeline across all datasets to ensure consistency. Student sequences with fewer than 20 interactions were removed to provide sufficient context for modeling and evaluation. All outcome labels were binarized based on dataset-specific scoring policies, consistent with prior work. Data were split at the student level: 90\% of learners were used for training, and 10\% were held out for testing, preventing data leakage across temporal sequences. For KUL-Rec, training was conducted using sliding windows of 10 interactions to simulate real-time, memory-driven updates. Baseline models were trained using the same filtered and chronologically ordered sequences.

\subsubsection{Baseline Models and Evaluation}

We compare KUL-Rec against eight established models representing major architectural paradigms in educational modeling:

\begin{itemize}
    \item \textbf{Sequence models:} Deep Knowledge Tracing (DKT) using an LSTM \citep{piech2015deep}.
    \item \textbf{Memory-augmented models:} DKVMN \citep{zhang2017dynamic} and IEKT \citep{long2021tracing}, which incorporate external key–value or interpretive memory components.
    \item \textbf{Transformer-based models:} AKT \citep{ghosh2020context}, a widely adopted model for concept-level interaction data.
    \item \textbf{Graph-based models:} GKT-MHA and GKT-PAM \citep{nakagawa2019graph}, which incorporate prerequisite relations through attention-based mechanisms.
    \item \textbf{State-space models:} Mamba4KT \citep{cao2024mamba4kt} and DKT2 using xLSTM \citep{zhou2025revisiting}, which replace attention mechanisms with linear-time memory updates.
\end{itemize}

All baseline models were implemented using publicly available code or reimplemented from the original papers when code was not accessible. Hyperparameters were kept consistent with those recommended in each source and optimized using the Adam optimizer (learning rate = 0.001, batch size = 256). Additional implementation details are provided in Appendix~\ref{appendix:datasets}.

We do not include comparisons with large language models (LLMs), as they require extensive prompt engineering and are not currently suited for continual, interpretable adaptation in real-time educational settings \citep{mazzullo2023learning, chuang2023evolving}. Instead, we focus on models with architectural transparency and compatibility with sparse supervision.
All experimental assets were used in accordance with their respective academic licenses (Appendix~\ref{appendix:datasets}).

\subsubsection{Bidirectional Ranking: Identifying the Learning Frontier}

Traditional models rank predicted success probabilities in descending order, assuming higher probabilities always indicate better recommendations. However, from an instructional perspective, both high and low predictions carry meaning. High-confidence predictions may indicate mastered concepts, while low-confidence ones may signal that the content is still too advanced. Neither is optimal for learning.

Instead, effective exercise recommendation aims to identify the \emph{learning frontier}—content that is appropriately challenging but still within reach. To evaluate whether the model identifies this frontier, we introduce a bidirectional ranking procedure. If the learner’s response to the recommended item is correct (\( r_{t+1} = 1 \)), the correct ranking places the item near the top. If the response is incorrect (\( r_{t+1} = 0 \)), the ideal ranking would have placed the item near the bottom, suggesting it was appropriately identified as difficult.

This criterion operationalizes the learning frontier by rewarding high confidence on attainable items and low confidence on items beyond reach, and it standardizes offline evaluation on KT logs where only one next item is observed.

\subsubsection{Rank-Sensitive Evaluation Metrics}
\label{sec:metrics}

To quantify how well the system identifies appropriate next exercises, we adopt rank-based metrics that reflect the pedagogical value of each recommendation. Specifically, we compute Recall@K and nDCG for each learner interaction.

Recall@K measures whether the actual item attempted appeared in the top \( K \) ranked items under the direction determined by the observed outcome:
\[
\text{Recall@}K_t = \mathbb{1}\!\left[q_{t+1} \in \text{TopK}(\pi_t^{r_{t+1}})\right],
\]
where \( \pi_t^{r_{t+1}} \) denotes the ranking direction: descending if the response was correct, ascending if incorrect.

nDCG@K assigns higher weight to items ranked closer to the ideal position:
\[
\text{nDCG} = \frac{1}{\log_2(1 + \text{rank}(q_{t+1}))},
\]
where \( y_{t+1} = r_{t+1} \) and the item is considered more relevant if it aligns with the learner’s demonstrated ability.

These metrics emphasize ranking quality over classification accuracy, aligning model evaluation with the core goal of personalized, effective recommendation.

\subsubsection{Online training loop}
\label{sec:online_loop}

We instantiate the procedure in Section~\ref{sec:train_adapt} with a fixed window length \(W=10\). For tabular benchmarks, each window is optimized with binary cross-entropy using Adam (learning rate \(1.5\times 10^{-2}\)). We run \(E=10\) local update epochs per window, writing \((\mathbf{h}_1, \mathbf{z}_2^{\ast})\) to memory after each epoch. Model weights persist across windows and are never reset.

The consolidation phase uses 256 sparse seeds evolved for \(U=8\) recurrent steps under \(\mathbf{J}\). Recalled traces are projected back to student space and used to fine-tune the slow learner for up to 500 iterations with early stopping after 100 non-improving updates. Unless stated otherwise, decay and plasticity are set to \(\rho=0.5\) and \(\eta=0.5\). These settings are held constant across benchmark evaluations to ensure comparability. Ablations vary the presence of the CLS-style memory and the LIT mechanism to isolate their contributions.

\subsubsection{Investigating Learning Dynamics}

To explore how neuro-inspired mechanisms affect model learning, we studied the interaction between plasticity, forgetting, and consolidation in our model. We conducted controlled experiments on \textsc{ASSIST2009} (mathematics) and \textsc{EdNet-Sm} (English), selected for their moderate size and runtime efficiency. Specifically, we varied the forgetting rate \(\rho \in \{0.1, 0.5, 0.9\}\), the plasticity parameter \(\eta \in \{0.1, 0.5, 0.9\}\), the number of local Hebbian updates per mini-batch \(\in \{10, 100\}\), and the early stopping patience for replay \(\in \{10, 100\}\).

\subsubsection{Results}
\label{parameters}

\begin{figure}[!h]
    \centering
    \includegraphics[width=1.0\linewidth]{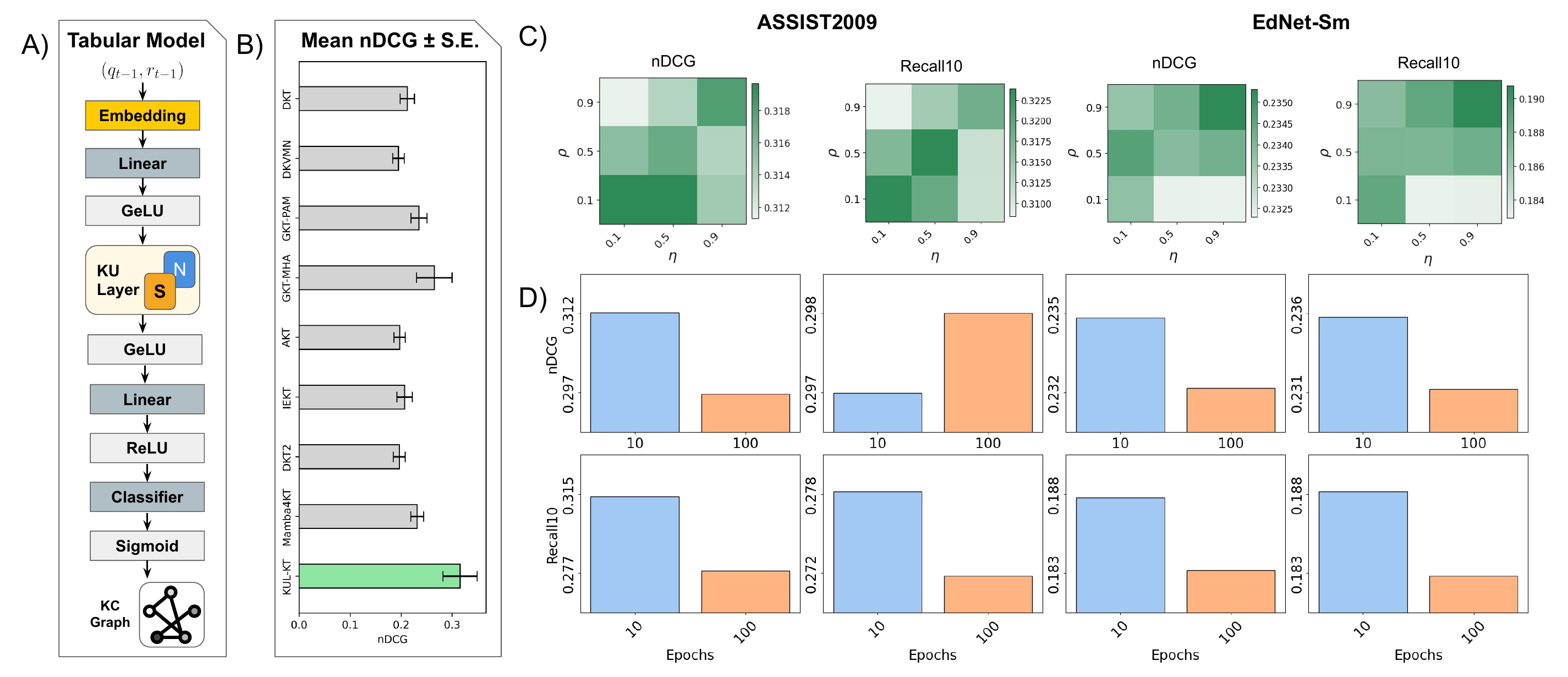}
    \caption{\textbf{Tabular dataset results.} 
    \textbf{A)} Model architecture of KUL-Rec used in this experiment. 
    \textbf{B)} Histogram of nDCG scores across all ten datasets, showing consistent performance gains over recurrent, transformer, and graph-based baselines. \newline    
    \textbf{Plasticity, retention, and consolidation in continual learning.} 
    \textbf{C)} Heatmaps of nDCG and Recall@10 across combinations of encoding rate (\(\eta\)) and forgetting rate (\(\rho\)) on \textsc{ASSIST2009} and \textsc{EdNet-Sm}. On math tasks, lower \(\eta\) and moderate forgetting (\(\rho\)) perform best, while language tasks benefit from higher values. Performance is relatively stable across settings. 
    \textbf{D)} Effect of replay consolidation iterations. Fewer replay steps yield better generalization, supporting the view that lighter rehearsal avoids overfitting. Small improvements in nDCG result from increased patience and extended consolidation epochs.
    }
    
    \label{fig:tabular_data_res}
\end{figure}

Table~\ref{tab:avg_results} reports macro-averaged rank performance across ten publicly available learner interaction datasets, using four complementary metrics: nDCG, Recall@1, Recall@5, and Recall@10. Figure~\ref{fig:tabular_data_res} summarizes the architecture and benchmark distribution. These metrics reflect different aspects of recommendation quality. While nDCG accounts for both relevance and position in the ranked list, Recall@K measures whether the correct next item appears within the top \(K\) predictions. Recall@1 captures high-precision targeting (e.g., ideal for fixed-length quizzes), while Recall@5 and @10 assess broader coverage—relevant for exploratory learning or scaffolded difficulty.

KUL-Rec consistently outperforms all baseline models across all four metrics. Compared to the strongest competing model (GKT-MHA), KUL-Rec improves nDCG by +0.051 (0.316 vs. 0.265; +19.2\% relative) and Recall@10 by +0.094 (0.305 vs. 0.211; +44.6\% relative). It also achieves a Recall@1 score of 0.123, nearly triple the baseline average, highlighting its strength in precise next-item recommendation. These improvements generalize across diverse datasets varying in subject domain, learner population size, and interaction sparsity. Full per-dataset scores and standard errors are reported in Appendix~\ref{app:tables}.

\begin{table}[!h]
\centering
\caption{Macro-averaged rank performance across ten tabular datasets ($\pm$\,standard error).}
\label{tab:avg_results}
\small
\begin{tabular}{lcccc}
\toprule
Model & nDCG $\uparrow$ & Recall@1 $\uparrow$ & Recall@5 $\uparrow$ & Recall@10 $\uparrow$\\
\midrule
DKT          & 0.212 ± 0.014 & 0.014 $\pm$ 0.006 & 0.052 $\pm$ 0.016 & 0.108 ± 0.027\\
DKVMN        & 0.195 ± 0.011 & 0.004 $\pm$ 0.002 & 0.026 $\pm$ 0.010 & 0.064 ± 0.019\\
GKT-PAM      & 0.235 ± 0.016 & 0.023 $\pm$ 0.007 & 0.091 $\pm$ 0.020 & 0.172 ± 0.035\\
GKT-MHA      & 0.265 ± 0.035 & 0.058 $\pm$ 0.029 & 0.139 $\pm$ 0.049 & 0.211 ± 0.060\\
AKT          & 0.197 ± 0.011 & 0.006 $\pm$ 0.002 & 0.033 $\pm$ 0.010 & 0.076 ± 0.019\\
IEKT         & 0.207 ± 0.015 & 0.011 $\pm$ 0.005 & 0.054 $\pm$ 0.019 & 0.107 ± 0.031\\
DKT2         & 0.196 ± 0.012 & 0.005 $\pm$ 0.002 & 0.033 $\pm$ 0.012 & 0.074 ± 0.024\\
Mamba4KT     & 0.231 ± 0.013 & 0.018 $\pm$ 0.005 & 0.091 $\pm$ 0.018 & 0.169 ± 0.027\\
\textbf{KUL-Rec} & \textbf{0.316 ± 0.034} & \textbf{0.123 $\pm$ 0.037} & \textbf{0.244 $\pm$ 0.045} & \textbf{0.305 ± 0.044}\\
\bottomrule
\end{tabular}
\end{table}

\noindent
All source code, configuration files, and trained models used in this experiment are available at: \url{https://github.com/mims-harvard/KUL-Rec}.

\subsubsection{Plasticity, Retention, and Consolidation}
\label{sec:plasticity}

To better understand how KUL-Rec supports continual learning, we investigate the effects of three biologically inspired learning parameters: encoding rate, memory retention, and consolidation. These parameters are grounded in complementary learning systems (CLS) theory and correspond to core phases of human learning: plasticity (encoding new experiences), forgetting (retaining or discarding information), and consolidation (replaying and integrating past experiences). Full metric tables are provided in Appendix~\ref{app:ablation_tables}.

We conduct these experiments on two pedagogically distinct datasets. \textsc{ASSIST2009} focuses on mathematics, where concepts are revisited frequently and reinforcement is common. In contrast, \textsc{EdNet-Sm} is drawn from an English second-language tutoring platform, where learning sequences are more varied and concept repetition is less predictable.

\paragraph{Encoding rate.}
We vary the encoding rate (\(\eta\)) to test how quickly the model incorporates new information. On \textsc{ASSIST2009}, a lower encoding rate (\(\eta = 0.1\)) leads to better performance, suggesting that gradual learning stabilizes internal representations when the same concepts are encountered repeatedly. On \textsc{EdNet-Sm}, a higher encoding rate (\(\eta = 0.9\)) is more effective, supporting the idea that fast-changing linguistic inputs require more rapid adaptation.

\paragraph{Retention (forgetting rate).}
We also explore how retention, governed by the forgetting parameter (\(\rho\)), interacts with content structure. Moderate forgetting (\(\rho = 0.5\)) performs best in the structured environment of \textsc{ASSIST2009}, while high retention (\(\rho = 0.9\)) is better suited to the more diverse sequences in \textsc{EdNet-Sm}. These findings align with cognitive studies suggesting that optimal retention varies with the frequency and regularity of concept exposure \citep{antony2017retrieval}.

\paragraph{Magnitude of effects.}
Although both parameters influence performance, the overall variation is modest. Across all tested combinations, the change in nDCG and Recall@10 remains under 0.01 (Fig.~\ref{fig:tabular_data_res}C), suggesting that KUL-Rec is robust to hyperparameter selection and supports stable learning across a range of conditions.

\paragraph{Consolidation iterations.}
Finally, we assess the role of memory replay during consolidation by varying the number of local update iterations per batch (10 vs. 100). Fewer replay iterations (10) consistently lead to higher performance across both datasets (Fig.~\ref{fig:tabular_data_res}D), indicating that excessive replay can cause overfitting to short-term noise. Increasing early stopping patience yields a small but consistent improvement in nDCG (\(\Delta \text{nDCG} \approx 0.01\)), and increasing the total number of consolidation epochs adds an additional minor gain (\(\Delta \text{nDCG} \approx 0.02\)). These findings support the pedagogical intuition that brief, focused review of recent material is more effective than prolonged rehearsal.


\subsubsection{Computational Efficiency and Resource Requirements}
\label{sec:runtime}

We evaluate KUL-Rec’s computational efficiency with respect to training time, inference latency, and memory usage—three factors that directly impact its applicability in real-world educational environments, particularly where resources or responsiveness are constrained.

Table~\ref{tab:efficiency_assist2009} presents comparative results on the \textsc{ASSIST2009} dataset. KUL-Rec achieves a favorable trade-off between speed and memory usage, with competitive training time (2.85\,s per epoch) and significantly lower inference latency (0.03\,s) compared to strong baselines such as Mamba4KT (2.23\,s) and GKT-MHA (4.72\,s). Its peak GPU memory usage is 0.09\,GB—substantially lower than both Mamba4KT (0.62\,GB) and GKT-MHA (7.98\,GB).

These efficiency gains are the result of KUL-Rec’s biologically inspired architecture. The model avoids backpropagation through time and does not require caching of hidden states, attention stacks, or past activations. Instead, it relies on sparse, local Hebbian updates and associative memory replay, yielding constant memory and compute per interaction. This allows the model to scale gracefully to longer sequences without increasing inference time or memory consumption.

Internally, KUL-Rec’s associative memory consists of five matrices: a recurrent connectivity matrix \( \mathbf{J} \in \mathbb{R}^{M \times M} \), and four projection matrices \( \mathbf{U}, \mathbf{V} \in \mathbb{R}^{M \times d} \) linking memory to the input and output spaces. With \( M = 2048 \) memory units and \( d = 64 \) dimensions, the total number of parameters remains under 5 million, corresponding to a static model size of 17.29\,MB (in FP32). While this size exceeds that of some lightweight architectures, its runtime memory and latency profile are significantly more stable.

In contrast, transformer and graph-based models must store and access per-step representations, such as key–value matrices or node embeddings, which grow with sequence length. This scaling leads to substantial increases in memory footprint and latency, especially during inference. Even optimized recurrent architectures such as Mamba4KT retain temporal dependencies that introduce overhead. KUL-Rec’s constant-time associative recall makes it especially well suited for deployment in time-sensitive or resource-limited educational systems, such as real-time tutoring, mobile learning apps, or classroom-based interventions.

\begin{table}[!h]
\centering
\caption{Computational efficiency on \textsc{ASSIST2009} (single NVIDIA A100, 40GB). We report per-epoch training time, end-to-end inference latency, peak GPU memory (VRAM), and static model size.}
\label{tab:efficiency_assist2009}
\small
\resizebox{0.98\linewidth}{!}{
\begin{tabular}{lcccc}
\toprule
\textbf{Model} & \textbf{Training Time (s)} & \textbf{Inference Time (s)} & \textbf{VRAM (GB)} & \textbf{Model Size (MB)}\\
\midrule
GKT-MHA     & 4.90 & 4.72 & 7.98 & 0.15 \\
Mamba4KT    & \textbf{2.79} & 2.23 & 0.62 & 2.23 \\
\textbf{KUL-Rec} & 2.85 & \textbf{0.03} & \textbf{0.09} & 17.29 \\
\bottomrule
\end{tabular}
}
\end{table}

\subsection{Short Answer Data Experiment}
\label{sec:shortansdata}

To evaluate KUL-Rec in a real-world instructional setting, we deployed the model in a 13-week graduate-level course and used it to generate personalized short-answer quizzes. This experiment tested the model's capacity to function in a low-data, open-response environment and to deliver value both algorithmically and pedagogically. We assessed: (i) the quality of exercise recommendation through ranking metrics computed on candidate question banks, and (ii) learner experience through weekly surveys of perceived difficulty and helpfulness.

\begin{figure}[h]
    \centering
    \includegraphics[width=1\linewidth]{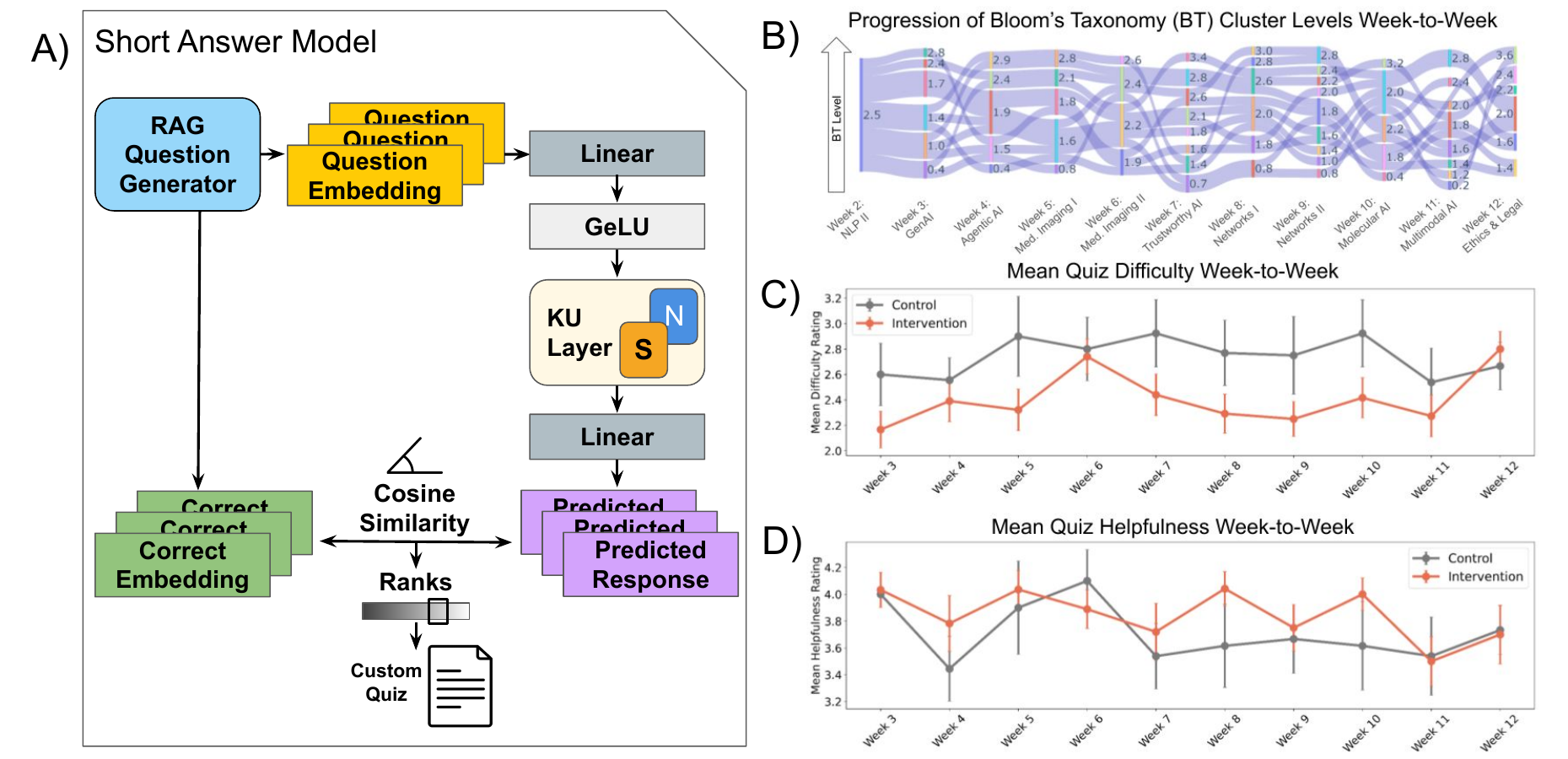}
    \caption{\textbf{Short-answer adaptation pipeline and outcomes.}
    \textbf{A:} Weekly personalized quiz generation. A Retrieval-Augmented Generation (RAG) pipeline constructs a candidate bank from learning objectives, readings, and Bloom's levels. Each student's KUL-Rec model ranks questions by predicted answer similarity, and one per objective is selected at the 66th percentile. 
    \textbf{B:} Sankey diagram showing the distribution of assigned Bloom's Taxonomy (BT) levels per student over 10 weeks, highlighting divergence in personalized trajectories.
    \textbf{C–D:} Mean ± s.e.m. survey ratings of perceived difficulty and helpfulness, showing significant improvement in the adaptive quiz condition.
    }
    \label{fig:shortanswer_results}
\end{figure}

\subsubsection{System Design and Personalized Quiz Construction}

To support personalized short-answer assessment, we combined a Retrieval-Augmented Generation (RAG) pipeline with weekly fine-tuning of individual KUL-Rec models. Each week, students submitted five short-answer responses to course readings. These were embedded using OpenAI’s \texttt{text-embedding-ada-002} model and used to update the student’s personalized instance of KUL-Rec via mean squared error loss.

Quiz question banks were generated using a RAG pipeline that integrated course learning objectives, weekly readings, and Bloom’s Taxonomy (BT), a framework for categorizing educational goals across six cognitive levels from recall to synthesis \citep{krathwohl2002revision}. For each learning objective and BT level combination, the pipeline retrieved semantically relevant content from the readings and prompted a large language model to generate a corresponding question–answer pair. Approximately 72 candidate questions were generated per week in total. The full pipeline, including indexing, retrieval, prompt scaffolding, and generation orchestration, is detailed in Appendix~\ref{sec:rag_pipeline}.

To rank candidate questions, KUL-Rec predicted an embedding of the learner’s expected answer and computed cosine similarity with the correct answer embedding. One question per objective was then selected at the 66th percentile of this ranking to approximate the learner’s ZPD\citep{shabani2010vygotsky}. This quiz generation workflow is illustrated in Figure~\ref{fig:shortanswer_results}A.

We also examined how different embedding models affect the short-answer ranking process. While performance varied across models, we found that the overall ranking quality and question selection were robust. In particular, the system’s ability to identify conceptually appropriate questions held steady across embedding spaces, even when surface-level similarity varied. Although our deployed model (\texttt{ada-002}) showed weaker correlation with student-written answers than newer alternatives such as \texttt{bge}, the top-ranked items remained consistent in meaning and pedagogical intent. This suggests that our use of semantic embeddings as flexible representations of meaning rather than strict templates, supports fair and stable recommendation. Additional details, including alignment metrics and threshold evaluations, are provided in Appendix~\ref{app:embedding_robustness}.

\subsubsection{Classroom Deployment and Learner Experience}

The study was conducted over 13 weeks in a graduate-level course with 38 enrolled MSc and PhD students. Each week, students could opt in to receive a personalized quiz generated by KUL-Rec or a static control quiz composed of instructor-selected questions. Quizzes were delivered through the university’s Canvas Learning Management System (LMS), and distribution was automated via the Canvas API. Each quiz contributed 1\% to the course grade, awarded for completion regardless of correctness.

Participation in the adaptive condition was entirely voluntary. Students who opted in received quiz questions selected at their predicted ZPD level. Those who did not received control quizzes with an average BT level of 2.5. All data were anonymized prior to analysis, and no personally identifiable information was collected or stored. The study was conducted under Institutional Review Board (IRB) approval, with exemption granted under 45 CFR 46.104(d)(1)(2). Consent forms were provided, and signature requirements were waived for students using the system. Further details on IRB protocols, participant instructions, and survey instrumentation are provided in Appendix~\ref{appendix:human_subjects}.

Before each quiz, students were informed that the assessment was intended to support learning, and they were encouraged to review the course readings. After each quiz, students completed a short survey evaluating how much time they spent studying, how difficult they found the quiz, and how helpful it was for consolidating key ideas. These self-report data provide insight into student perceptions of the system’s instructional value.

\subsubsection{Individual Model Training and Deployment}

Each student received a personalized instance of KUL-Rec updated weekly with their latest short-answer responses. Model training was conducted on an NVIDIA A100 (40 GB), with adaptation time averaging 1–2 minutes per student per week. At inference, the model predicted the likely student response to each candidate question, then selected questions at the ZPD-aligned percentile for quiz delivery. All model instances were managed in parallel, and quiz recommendations were generated automatically through a browser-based interface. The system’s source code is available at \url{https://github.com/gkuling/QuizGen-RAG}, and the course website is archived at \url{https://zitniklab.hms.harvard.edu/AIM2/}.

\subsubsection{Learner-Perceived Value}

To evaluate the instructional impact of the adaptive quizzes, we collected weekly survey data from students using a 5-point Likert scale. Students rated the perceived difficulty and helpfulness of the quizzes they received. Those in the adaptive condition (approximately 20 participants per week) reported significantly lower difficulty ratings ($2.40 \pm 0.77$) and higher helpfulness ratings ($3.86 \pm 0.83$) than peers in the static quiz condition (difficulty = $2.75 \pm 0.86$, helpfulness = $3.69 \pm 0.92$), as shown in Table~\ref{tab:survey_results}. All differences were statistically significant ($p < 0.05$, Wilcoxon signed-rank test), indicating that personalized quizzes were not only better aligned with each learner’s current ability, but also perceived as more useful.

\begin{table}[h!]
    \caption{Average survey ratings across the semester. Ratings use a 5-point Likert scale (lower is better for difficulty, higher is better for helpfulness).}
    \label{tab:survey_results}
    \centering
    \begin{tabular}{lcc}
    \toprule
    \textbf{Metric} & \textbf{Adaptive} & \textbf{Static} \\
    \midrule
    Difficulty ↓ & $2.40\pm0.77$ & $2.75\pm0.86$ \\
    Helpfulness ↑ & $3.86\pm0.83$ & $3.69\pm0.92$ \\
    \bottomrule
    \end{tabular}
    
\end{table}

Figure~\ref{fig:shortanswer_results}B visualizes how the system adapted over time. A Sankey diagram shows the average Bloom level of assigned questions per student over 10 weeks. Diverging paths reflect how the model learned to tailor difficulty, assigning more advanced cognitive tasks as student mastery improved. This suggests that KUL-Rec was able to track individual learning progress and adjust recommendations accordingly, supporting differentiated instruction in a scalable manner.

\section{Discussion}
\label{sec:discussion}

This study introduces KUL-Rec, a biologically inspired system for personalized exercise recommendation, designed to support scalable, adaptive learning across both structured and open-ended domains. Drawing on complementary learning systems theory \cite{mcclelland1995there}, KUL-Rec integrates fast Hebbian encoding with slower replay-based consolidation to enable continual, individualized adaptation. Across two settings—a benchmarked corpus of tabular datasets and a graduate-level classroom deployment—the model consistently delivered rank-based improvements in recommending pedagogically appropriate exercises.

Our findings support three key conclusions relevant to educational researchers and practitioners. Reframing next-item prediction as exercise recommendation aligns more closely with instructional goals, particularly when personalized practice and formative assessment are the objective. Evaluating systems based on rank-sensitive metrics such as nDCG and Recall@K better captures whether learners are receiving exercises at the right level of challenge, rather than simply predicting correctness. The integration of fast encoding, forgetting, and bounded replay supports personalization in low-data, real-time settings. KUL-Rec performs well even when updated with only a handful of learner interactions each week, offering a lightweight and scalable solution for formative feedback, targeted retrieval, or adaptive quiz delivery. Its ability to operate on vectorized inputs, rather than relying on predefined concept vocabularies, allows for flexible use across modalities—including short-answer responses—while maintaining a consistent computational footprint. Embedding-based personalization introduces new opportunities and responsibilities. Our embedding robustness analysis shows that the choice of semantic encoder can impact system behavior, especially in open-ended assessment. While KUL-Rec’s threshold-based selection remains stable across models, embedding audits should become a routine part of system validation. As educational AI increasingly incorporates open-response generation and evaluation, fairness, interpretability, and linguistic variability must be addressed explicitly.

Several limitations point directly to next steps in development and evaluation. The current design trains a separate KUL-Rec instance per learner, which may be impractical at MOOC scale, so future work should explore parameter sharing, lightweight per-learner adapters, model distillation, and federated aggregation to retain personalization while reducing cost. Our classroom study involved a small, self-selected graduate cohort, so broader evaluations across K–12 and diverse postsecondary settings, ideally with preregistered protocols and sufficient statistical power, are needed to assess generalizability and equity. Although students rated adaptive quizzes as helpful and well aligned, the system offers limited transparency about why specific items were recommended, suggesting development of per-recommendation rationales, summaries of influential prior responses, and teacher dashboards that support inspection and override. Because embedding choice can introduce demographic or cultural bias, especially in open-ended text, future iterations should include routine embedding audits, multilingual encoders, calibration procedures, and subgroup fairness checks with ongoing monitoring. More generally, embedding-based personalization is promising for continual, low-data adaptation, and its deployment should be paired with validation that includes learning outcomes, persistence, and engagement, not only rank metrics, along with online A/B tests in the LMS to reduce exposure bias and approximate counterfactual evaluation.

KUL-Rec demonstrates how biologically inspired memory mechanisms can be translated into scalable educational tools that support adaptive assessment. Its integration with standard learning platforms, low memory and latency requirements, and flexibility across question types make it a practical candidate for deployment in real classrooms. The model supports both the delivery of personalized quizzes and the design of learning trajectories that stay aligned with a student’s current level of understanding—key goals in scaffolding, formative feedback, and mastery learning. Future development should explore federated adaptation, instructor oversight tools, and richer response modeling to support open-ended feedback. As adaptive systems become more integrated into instruction, transparency, equity, and instructional alignment must guide design choices.

\section{Conclusion}

KUL-Rec offers an efficient and extensible approach to personalized learning. By combining Hebbian-style memory with rank-based evaluation and embedding-aware supervision, it bridges the gap between scalable AI and pedagogical relevance. Its performance on tabular and short-answer tasks, as well as its deployment in a real course setting, suggests that biologically grounded, rank-sensitive learning systems can contribute meaningfully to the next generation of personalized educational technologies.

\section*{Acknowledgments}

We are deeply grateful to the students in AI in Medicine II (AIM2) and BMI 702 at Harvard Medical School who participated in this study. Your thoughtful engagement and willingness to interact with an experimental AI-powered learning system made this research possible.

We thank Teaching Fellows Yasha Ektefaie, Yepeng Huang, and Courtney A. Shearer for their instructional support and collaboration. Special thanks to Ruthie Johnson and Michelle Li for manuscript proofreading and feedback.

G.K. is supported by the Curriculum Fellows Program at Harvard Medical School. This work was supported in part by NIH R01-HD108794, NSF CAREER 2339524, and US DoD FA8702-15-D-0001, as well as awards from the Harvard Data Science Initiative, Amazon Faculty Research, Google Research Scholar Program, AstraZeneca Research, Roche Alliance with Distinguished Scientists, Sanofi iDEA-iTECH, Pfizer Research, Chan Zuckerberg Initiative, John and Virginia Kaneb Fellowship at Harvard Medical School, Biswas Computational Biology Initiative (Milken Institute), Harvard Medical School Dean's Innovation Fund for the Use of Artificial Intelligence, and the Kempner Institute for the Study of Natural and Artificial Intelligence at Harvard University.

This study was reviewed and approved by the Harvard Faculty of Medicine Institutional Review Board (IRB24-1600), with exempt status under 45 CFR 46.104(d)(1)(2). Additional ethics and consent details are provided in Appendix~\ref{appendix:human_subjects}.

Any opinions, findings, or conclusions expressed in this material are those of the authors and do not necessarily reflect the views of the supporting institutions.

\bibliographystyle{unsrt}  
\bibliography{references}  

\newpage
\appendix
\newpage
\section{Dataset Statistics and Preprocessing}
\label{appendix:datasets}

This appendix provides an overview of the benchmark datasets used to evaluate KUL-Rec and all baseline models. These datasets span a diverse range of student populations, domains, and interaction styles, making them well-suited for evaluating model generalizability in knowledge tracing.

\subsection{Licenses and Attribution for External Assets}

This work makes use of the following publicly available datasets and codebases. All assets are properly cited in the main text, and their respective licenses and terms of use were reviewed and adhered to in accordance with the NeurIPS Code of Ethics.

\vspace{1em}
\noindent\textbf{Public Datasets Used}
\begin{itemize}
    \item \textsc{ASSISTments} (2009, 2012, 2015, 2017): 
    \url{https://sites.google.com/view/assistmentsdatamining/}
    — CC BY 4.0 License.
    \item \textsc{EdNet} (small, large): 
    \url{https://github.com/riiid/ednet} 
    — Custom academic license, permitted for research use.
    \item \textsc{XES3G5M}: 
    \url{https://github.com/ai4ed/XES3G5M}
    — MIT License.
    \item \textsc{Bridge to Algebra 2006, Algebra 2005}:
    \url{https://pslcdatashop.web.cmu.edu/}
    — Licensed for academic research use via the DataShop platform.
    \item \textsc{NIPS34 diagnostic dataset}:
    \url{https://eedi.com/projects/neurips-education-challenge}
    — CC BY-NC-ND 4.0.
\end{itemize}

\vspace{1em}
\noindent\textbf{Baseline Model Codebases}
\begin{itemize}
    \item Deep Knowledge Tracing \textbf{DKT}:
    \url{https://github.com/chrispiech/DeepKnowledgeTracing}
    — MIT License.
    \item Dynamic Key-Value Memory Networks \textbf{DKVMN}:
    \url{https://github.com/jennyzhang0215/DKVMN}
    — MIT License.
    \item Individual Estimation Knowledge Tracing \textbf{IEKT}:
    \url{https://github.com/ApexEDM/iekt}
    — Apache 2.0 License.
    \item Attention Knowledge Tracing \textbf{AKT}:
    \url{https://github.com/arghosh/AKT}
    — MIT License.
    \item Graph  Knowledge Tracing\textbf{GKT-MHA / GKT-PAM}:
    \url{https://github.com/jhljx/GKT}
    — MIT License.
    \item \textbf{Mamba4KT}:
No official public implementation or license was available at the time of writing. We implemented the model from scratch based on the original paper \cite{cao2024mamba4kt} to ensure a fair comparison. Our reimplementation will be released under an open-source license upon publication.
    \item Deep Knowledge Tracing 2 \textbf{DKT2 (xLSTM)}:
    \url{https://github.com/codebase-2025/DKT2}
    Although a GitHub repository exists for DKT2, no valid license was specified at the time of submission. We implemented the model from scratch based on the original paper \cite{zhou2025revisiting}. Our implementation was used solely for benchmarking and will be released under an open-source license upon publication.

\end{itemize}

All assets were used in accordance with their respective terms and solely for non-commercial academic research purposes. No modifications were made to the datasets themselves beyond standard preprocessing (e.g., filtering, chronological sorting) for modeling purposes.

\newpage
\section{Full Results Tables}
\label{app:tables}

In the main text, we reported macro-averaged scores across ten public KT benchmarks to summarize overall performance trends. Here, we expand those results with dataset-specific scores for four rank-based metrics: nDCG, Recall@1, Recall@5, and Recall@10. These tables provide a more detailed view of KUL-Rec’s performance across diverse datasets varying in domain, size, and interaction sparsity.

\begin{table}[h!]
\centering
\caption{nDCG results across all datasets (best per dataset in \textbf{bold}).}
\label{ndcg_results}
\small
\resizebox{\textwidth}{!}{%
\begin{tabular}{lrrrrrrrrrrr}
\toprule
\textbf{Dataset} & \textbf{Algebra2005} & \textbf{Bridge2006} & \textbf{ASSIST2009} & \textbf{ASSIST2012} & \textbf{ASSIST2015} & \textbf{ASSIST2017} & \textbf{EdNet-Sm} & \textbf{EdNet-LG} & \textbf{NIPS34} & \textbf{XES3G5M} & \textbf{Average $\pm$ SE} \\
\midrule
DKT       & 0.252 & 0.204 & 0.279 & 0.161 & 0.258 & 0.206 & 0.194 & 0.189 & 0.238 & 0.138 & 0.212 $\pm$ 0.014 \\
DKVMN     & 0.219 & 0.199 & 0.200 & 0.157 & 0.254 & 0.190 & 0.193 & 0.190 & 0.222 & 0.123 & 0.195 $\pm$ 0.011 \\
GKT-PAM   & 0.234 & 0.293 & 0.280 & 0.191 & \textbf{0.317} & 0.226 & 0.197 & 0.196 & 0.252 & 0.164 & 0.235 $\pm$ 0.016 \\
GKT-MHA   & 0.320 & 0.525 & \textbf{0.322} & 0.170 & 0.281 & 0.245 & 0.195 & 0.196 & 0.254 & 0.142 & 0.265 $\pm$ 0.035 \\
AKT       & 0.215 & 0.189 & 0.209 & 0.159 & 0.250 & 0.197 & 0.192 & 0.193 & 0.237 & 0.129 & 0.197 $\pm$ 0.011 \\
IEKT      & 0.210 & 0.307 & 0.205 & 0.165 & 0.229 & 0.209 & 0.187 & 0.189 & 0.241 & 0.126 & 0.207 $\pm$ 0.015 \\
DKT2      & 0.186 & 0.179 & 0.236 & 0.160 & 0.254 & 0.198 & 0.197 & 0.192 & 0.230 & 0.129 & 0.196 $\pm$ 0.012 \\
Mamba4KT  & 0.203 & 0.283 & 0.275 & 0.233 & 0.241 & 0.243 & 0.188 & \textbf{0.248} & 0.251 & 0.151 & 0.231 $\pm$ 0.013 \\
KUL-Rec   & \textbf{0.448} & \textbf{0.532} & 0.316 & \textbf{0.274} & 0.274 & \textbf{0.347} & \textbf{0.236} & 0.233 & \textbf{0.321} & \textbf{0.177} & \textbf{0.316 $\pm$ 0.034} \\
\bottomrule
\end{tabular}
}
\end{table}

\begin{table}[h!]
\centering
\caption{Recall@1 results across all datasets (best per dataset in \textbf{bold}).}
\label{recall11_results}
\small
\resizebox{\textwidth}{!}{%
\begin{tabular}{lrrrrrrrrrrr}
\toprule
           Model & Algebra2005 & Bridge2006 & ASSIST2009 & ASSIST2012 & ASSIST2015 & ASSIST2017 & EdNet-Sm & EdNet-LG & NIPS34 & XES3G5M &  Average $\pm$ SE \\
\midrule
             DKT &       0.019 &      0.000 &      0.058 &      0.000 &      0.032 &      0.005 &    0.005 &    0.003 &  0.007 &   0.008 & 0.014 $\pm$ 0.006 \\
           DKVMN &       0.005 &      0.000 &      0.001 &      0.000 &      0.026 &      0.001 &    0.005 &    0.003 &  0.002 &   0.001 & 0.004 $\pm$ 0.002 \\
         GKT-PAM &       0.017 &      0.020 &      0.057 &      0.010 &      0.067 &      0.014 &    0.006 &    0.006 &  0.015 &   \textbf{0.019} & 0.023 $\pm$ 0.007 \\
         GKT-MHA &       0.072 &      0.306 &      0.089 &      0.002 &      0.042 &      0.033 &    0.006 &    0.006 &  0.017 &   0.005 & 0.058 $\pm$ 0.029 \\
             AKT &       0.007 &      0.001 &      0.001 &      0.000 &      0.027 &      0.004 &    0.004 &    0.004 &  0.005 &   0.003 & 0.006 $\pm$ 0.002 \\
            IEKT &       0.014 &      0.058 &      0.004 &      0.001 &      0.019 &      0.006 &    0.003 &    0.003 &  0.006 &   0.001 & 0.011 $\pm$ 0.005 \\
            DKT2 &       0.001 &      0.000 &      0.015 &      0.000 &      0.019 &      0.002 &    0.007 &    0.003 &  0.003 &   0.003 & 0.005 $\pm$ 0.002 \\
        Mamba4KT &       0.008 &      0.049 &      0.040 &      0.006 &      0.019 &      0.014 &    0.003 &    0.025 &  0.012 &   0.001 & 0.018 $\pm$ 0.005 \\
\textbf{KUL-Rec} &       \textbf{0.275} &      \textbf{0.375} &      \textbf{0.110} &      \textbf{0.083} &      \textbf{0.073} &      \textbf{0.144} &    \textbf{0.034} &    \textbf{0.031} &  \textbf{0.082} &   \textbf{0.019} & \textbf{0.123 $\pm$ 0.037} \\
\bottomrule
\end{tabular}
}
\end{table}

\begin{table}[h!]
\centering
\caption{Recall@5 results across all datasets (best per dataset in \textbf{bold}).}
\label{recall5_results}
\small
\resizebox{\textwidth}{!}{%
\begin{tabular}{lrrrrrrrrrrr}
\toprule
           Model & Algebra2005 & Bridge2006 & ASSIST2009 & ASSIST2012 & ASSIST2015 & ASSIST2017 & EdNet-Sm & EdNet-LG & NIPS34 & XES3G5M &  Average $\pm$ SE \\
\midrule
             DKT &       0.117 &      0.005 &      0.144 &      0.004 &      0.101 &      0.032 &    0.023 &    0.021 &  0.059 &   0.019 & 0.052 $\pm$ 0.016 \\
           DKVMN &       0.054 &      0.003 &      0.016 &      0.002 &      0.103 &      0.008 &    0.026 &    0.022 &  0.022 &   0.004 & 0.026 $\pm$ 0.010 \\
         GKT-PAM &       0.084 &      0.221 &      0.153 &      0.042 &      0.136 &      0.074 &    0.033 &    0.028 &  0.091 &   0.048 & 0.091 $\pm$ 0.020 \\
         GKT-MHA &       0.242 &      0.513 &      0.219 &      0.009 &      0.128 &      0.104 &    0.031 &    0.027 &  0.098 &   0.018 & 0.139 $\pm$ 0.049 \\
             AKT &       0.043 &      0.004 &      0.032 &      0.003 &      0.104 &      0.022 &    0.026 &    0.025 &  0.061 &   0.009 & 0.033 $\pm$ 0.010 \\
            IEKT &       0.062 &      0.212 &      0.030 &      0.010 &      0.072 &      0.046 &    0.021 &    0.020 &  0.065 &   0.004 & 0.054 $\pm$ 0.019 \\
            DKT2 &       0.006 &      0.000 &      0.082 &      0.001 &      0.112 &      0.020 &    0.029 &    0.022 &  0.043 &   0.010 & 0.033 $\pm$ 0.012 \\
        Mamba4KT &       0.048 &      0.180 &      0.160 &      0.078 &      0.099 &      0.108 &    0.021 &    0.113 &  0.093 &   0.007 & 0.091 $\pm$ 0.018 \\
\textbf{KUL-Rec} &       \textbf{0.431} &      \textbf{0.532} &      \textbf{0.254} &      \textbf{0.208} &      \textbf{0.171} &      \textbf{0.285} &    \textbf{0.120} &    \textbf{0.115} &  \textbf{0.249} &   \textbf{0.078} & \textbf{0.244 $\pm$ 0.045} \\
\bottomrule
\end{tabular}
}
\end{table}

\begin{table}[h!]
\centering
\caption{Recall@10 results across all datasets (best per dataset in \textbf{bold}).}
\label{recall10_results}
\small
\resizebox{\textwidth}{!}{%
\begin{tabular}{lrrrrrrrrrrr}
\toprule
\textbf{Dataset} & \textbf{Algebra2005} & \textbf{Bridge2006} & \textbf{ASSIST2009} & \textbf{ASSIST2012} & \textbf{ASSIST2015} & \textbf{ASSIST2017} & \textbf{EdNet-Sm} & \textbf{EdNet-LG} & \textbf{NIPS34} & \textbf{XES3G5M} & \textbf{Average $\pm$ SE} \\
\midrule
DKT       & 0.231 & 0.045 & 0.237 & 0.011 & 0.195 & 0.074 & 0.059 & 0.054 & 0.151 & 0.026 & 0.108 $\pm$ 0.027 \\
DKVMN     & 0.132 & 0.018 & 0.055 & 0.007 & 0.195 & 0.028 & 0.061 & 0.057 & 0.084 & 0.005 & 0.064 $\pm$ 0.019 \\
GKT-PAM   & 0.165 & 0.399 & 0.231 & 0.078 & \textbf{0.296} & 0.143 & 0.073 & 0.063 & 0.195 & 0.075 & 0.172 $\pm$ 0.035 \\
GKT-MHA   & 0.397 & \textbf{0.626} & 0.311 & 0.025 & 0.229 & 0.164 & 0.067 & 0.064 & 0.198 & 0.034 & 0.211 $\pm$ 0.060 \\
AKT       & 0.105 & 0.022 & 0.093 & 0.009 & 0.191 & 0.046 & 0.065 & 0.063 & 0.151 & 0.012 & 0.076 $\pm$ 0.019 \\
IEKT      & 0.105 & 0.353 & 0.079 & 0.027 & 0.129 & 0.100 & 0.050 & 0.054 & 0.168 & 0.008 & 0.107 $\pm$ 0.031 \\
DKT2      & 0.021 & 0.001 & 0.162 & 0.007 & 0.232 & 0.049 & 0.069 & 0.059 & 0.128 & 0.016 & 0.074 $\pm$ 0.024 \\
Mamba4KT  & 0.098 & 0.281 & 0.250 & 0.170 & 0.192 & 0.230 & 0.054 & \textbf{0.197} & 0.202 & 0.021 & 0.169 $\pm$ 0.027 \\
KUL-Rec   & \textbf{0.471} & 0.579 & \textbf{0.315} & \textbf{0.270} & 0.235 & \textbf{0.336} & \textbf{0.188} & 0.185 & \textbf{0.343} & \textbf{0.125} & \textbf{0.305 $\pm$ 0.044} \\
\bottomrule
\end{tabular}
}
\end{table}

\newpage
\section{Ablation studies}
\label{app:ablation_tables}

\subsection*{Hebbian Decay Parameters}
\begin{wraptable}{r}{0.5\textwidth}
  \centering
  \caption{Hebbian decay ablation results on ASSIST2009 vs.\ EdNet-Sm}
  \small
  \resizebox{0.95\linewidth}{!}{  

  \begin{tabular}{llccc|ccc}
    \toprule
    & & \multicolumn{3}{c}{\bfseries ASSIST2009} 
         & \multicolumn{3}{c}{\bfseries EdNet-Sm} \\
    {\bfseries Metric} 
      & $\rho\backslash\eta$ & 0.1 & 0.5 & 0.9 
      & 0.1 & 0.5 & 0.9 \\
    \midrule
    \multirow{3}{*}{nDCG}
      & 0.1 & 0.320 & 0.320 & 0.314 
            & 0.234 & 0.232 & 0.232 \\
      & 0.5 & 0.315 & 0.317 & 0.314 
            & 0.235 & 0.234 & 0.234 \\
      & 0.9 & 0.311 & 0.314 & 0.318 
            & 0.234 & 0.234 & 0.235 \\
    \addlinespace
    \multirow{3}{*}{Recall@10}
      & 0.1 & 0.324 & 0.319 & 0.311 
            & 0.189 & 0.183 & 0.183 \\
      & 0.5 & 0.317 & 0.324 & 0.311 
            & 0.188 & 0.187 & 0.188 \\
      & 0.9 & 0.308 & 0.314 & 0.318 
            & 0.187 & 0.189 & 0.191 \\
    \bottomrule
  \end{tabular}
  }
\end{wraptable}

This section reports detailed ablation results for the encoding rate (\( \eta \)) and forgetting rate (\( \rho \)) parameters introduced in our Hebbian memory update rule. We evaluate all \(3 \times 3\) combinations across two datasets: ASSIST2009 and EdNet-Sm. Metrics including nDCG and Recall@10. While performance variations are modest (< 0.01), these results demonstrate that the model is robust to a range of plasticity and retention settings. Notably, lower encoding rates and moderate forgetting generally yield more stable performance on repetitive tasks like ASSIST2009, while higher plasticity helps on more variable data like EdNet-Sm.

\begin{wraptable}{r}{0.5\textwidth}
  \centering
  \caption{Consolidation‐phase ablation (10 vs.\ 100 epochs, with/without early stopping) on ASSIST2009 vs.\ EdNet-Sm}
  \small
    \resizebox{0.95\linewidth}{!}{  

  \begin{tabular}{llcc|cc}
    \toprule
    & & \multicolumn{2}{c}{\bfseries ASSIST2009} 
         & \multicolumn{2}{c}{\bfseries EdNet-Sm} \\
    {\bfseries Metric} 
      & Patience\textbackslash Epochs & 10 & 100 
      & 10 & 100 \\
    \midrule
\multirow[t]{2}{*}{nDCG} & 10 & 0.312 & 0.297 & 0.235 & 0.232 \\
 & 100 & 0.297 & 0.298 & 0.236 & 0.231 \\
\cline{1-6}
\multirow[t]{2}{*}{Recall10} & 10 & 0.314 & 0.278 & 0.188 & 0.183 \\
 & 100 & 0.278 & 0.272 & 0.188 & 0.183 \\
    \bottomrule
  \end{tabular}
  }
\end{wraptable}

\subsection*{Consolidation Parameters}

We ablate the number of replay iterations and the early stopping patience used during memory consolidation. We compare 10 vs. 100 update epochs per replay phase, with early stopping patience values of 10 and 100. Results suggest that shorter consolidation is generally sufficient, with 10 epochs and moderate patience providing optimal performance. Excessive replay appears to slightly degrade performance, especially in terms of Recall@10, supporting our design choice for lightweight consolidation.

\subsection*{CLS and LIT Ablation}
\begin{wraptable}{r}{0.5\textwidth}
\centering
\caption{CLS Ablation results}
\small
  \resizebox{0.95\linewidth}{!}{  

\begin{tabular}{llll}
\toprule
 & Dataset & ASSIST2009 & EdNet-Sm \\
Metric & Experiment &  &  \\
\midrule
\multirow[t]{3}{*}{nDCG} & \textbf{Proposed Model} &\textbf{ 0.316} & \textbf{0.236} \\
 & No CLS  & 0.313 & 0.233 \\
 & No LIT & 0.314 & 0.236 \\
\cline{1-4}
\bottomrule
\end{tabular}
}

\end{wraptable}

We isolate the contributions of (i) the CLS-style memory replay and (ii) the loss-aligned internal target (LIT) mechanism for computing surrogate outputs. Removing either component leads to measurable performance drops in at least one of the metrics across both datasets. This confirms that both biological replay and our local surrogate target are essential for maintaining accuracy and consistency under continual learning.

\newpage
\section{Retrieval-Augmented Generation (RAG) Pipeline for Quiz Generation}
\label{sec:rag_pipeline}

To support the automated generation of open-ended quiz questions aligned with weekly learning objectives, we developed a modular Retrieval-Augmented Generation (RAG) pipeline. This system integrates course content, curriculum planning, and large language models (LLMs) to produce personalized question–answer (Q\&A) banks for use in formative assessments.

The pipeline is designed to be transparent, customizable, and scalable, allowing instructors or researchers to adapt it to different syllabi, knowledge domains, and cognitive levels. It is structured into four modular components: semantic indexing of course materials, prompt scaffolding based on Bloom's Taxonomy, curriculum-driven schedule control, and end-to-end quiz generation.

\subsection*{Indexing Course Materials}

The pipeline begins by processing assigned weekly readings (PDF documents), which are split into overlapping chunks of 1024 tokens with 128-token overlaps. Each chunk is embedded into a high-dimensional vector space using the Azure OpenAI \texttt{text-embedding-ada-002} model. The resulting embeddings are stored in a persistent vector database using the \texttt{VectorStoreIndex} interface. If an index already exists for a given reading set, it is reloaded to avoid redundant computation.

\subsection*{Curriculum-Aware Prompt Construction}

Each week’s learning objectives are defined in a course timetable module. For each objective, the system iterates through six cognitive levels of Bloom's Taxonomy: Remember, Understand, Apply, Analyze, Evaluate, and Create. Each objective–level pair is used to retrieve semantically relevant content from the course materials, forming the basis for question generation.

A composite prompt is created by combining (i) the objective, (ii) the Bloom level instruction, and (iii) the retrieved supporting text. This prompt is passed to a large language model, which returns a Q\&A pair formatted for export. All prompts are assembled using predefined templates to ensure consistency and pedagogical structure.

\subsection*{Quiz Generation and Export}

The final orchestration script cycles through every combination of learning objective, reference reading, and Bloom level. For each, it retrieves supporting content, constructs a prompt, and generates a Q\&A pair using the LLM. These pairs are optionally shuffled, filtered, and exported to CSV format for downstream use in adaptive or static quiz delivery. The modular architecture allows for easy tuning (e.g., targeting higher-order Bloom levels or specific themes) and facilitates reuse in future iterations or courses.

\subsection*{Algorithm Summary}

Algorithm~\ref{alg:rag_quiz} presents a high-level summary of the quiz generation process. Detailed implementation is available in the public repository.

\begin{algorithm}[H]
  \caption{RAG-based Quiz Generation Pipeline}
  \label{alg:rag_quiz}
  \begin{algorithmic}[1]
    \Procedure{BuildIndex}{pdf\_folder}
      \State Load and tokenize readings with overlap
      \State Embed text chunks with \texttt{ada-002}
      \State Store or reload vector index
    \EndProcedure

    \Procedure{GenerateQuiz}{week, n}
      \State For each (objective, reference, Bloom level):
      \State \quad Retrieve top-$k$ relevant passages
      \State \quad Assemble prompt using templates
      \State \quad Generate Q\&A pair using LLM
      \State Shuffle and export top-$n$ pairs
    \EndProcedure
  \end{algorithmic}
\end{algorithm}

\newpage
\section{Embedding Robustness and Fairness in Short-Answer Evaluation}
\label{app:embedding_robustness}

To evaluate how embedding model choice impacts semantic assessment in KUL-Rec’s short-answer pipeline, we conducted a robustness audit across three widely used sentence embedding backbones: \texttt{mpnet}, \texttt{ada-002}, and \texttt{bge}. While each of these models is trained to capture semantic similarity between texts, their behavior can vary substantially in educational settings where student responses may be lexically diverse or conceptually nuanced.

\subsection*{Alignment Between Predicted and Actual Responses}

For each learner and week, we computed two cosine similarity values:
\begin{itemize}
    \item \textbf{Predicted Similarity (\texttt{pred\_sim})}: The similarity between KUL-Rec’s predicted response and the canonical correct answer.
    \item \textbf{Actual Similarity (\texttt{act\_sim})}: The similarity between the student’s submitted response and the canonical answer.
\end{itemize}

We then assessed how well each embedding model preserved the relationship between \texttt{pred\_sim} and \texttt{act\_sim} using three metrics:
\begin{itemize}
    \item \textbf{Spearman’s \(\rho\)}: Measures rank correlation between predicted and actual similarities.
    \item \textbf{Mean Absolute Error (MAE)}: Captures the average absolute difference between predicted and actual similarities.
    \item \textbf{Median Absolute Deviation (MAD\(\,\Delta\))}: A robust measure of spread in the error distribution.
\end{itemize}

\begin{table}[h]
\centering
\caption{Alignment metrics across embedding backbones (higher \(\rho\) is better; lower MAE and MAD\(\,\Delta\) are better).}
\label{tab:embed_alignment}
\small
\begin{tabular}{lccc}
\toprule
\textbf{Embedding Model} & \textbf{Spearman \(\rho\)} & \textbf{MAE} & \textbf{MAD\(\,\Delta\)} \\
\midrule
mpnet   & 0.068 & 0.727 & 0.045 \\
ada-002 & 0.020 & 0.693 & 0.040 \\
\textbf{bge}    & \textbf{0.243} & \textbf{0.546} & 0.044 \\
\bottomrule
\end{tabular}
\end{table}

Among the three models, \texttt{bge} provided the strongest rank correlation and lowest error, indicating that it best preserved the semantic distinctions between correct and incorrect answers. The \texttt{ada-002} model, while less sensitive to semantic variation, was used in the live system due to its stability and integration constraints. These findings underscore that embedding model choice is not neutral, and can meaningfully affect downstream evaluation quality in educational applications.

\subsection*{Threshold-Based Accuracy for Top-Ranked Recommendations}

To assess the functional implications of these differences, we examined how often the system’s top-ranked question fell within an acceptable semantic proximity to the correct answer. Specifically, for each student, we asked whether their actual response to the top-ranked question exceeded a cosine similarity threshold \( \theta \in \{0.5, 0.6, 0.7, 0.8, 0.9\} \).

\begin{table}[h]
\centering
\caption{Proportion of top-ranked items where actual response similarity exceeds threshold \(\theta\).}
\label{tab:embed_thresholds}
\small
\begin{tabular}{lccc}
\toprule
\textbf{Threshold \(\theta\)} & \textbf{mpnet} & \textbf{ada-002} & \textbf{bge} \\
\midrule
0.5 & 1.00 & 1.00 & 1.00 \\
0.6 & 1.00 & 1.00 & 1.00 \\
0.7 & 0.967 & 0.933 & 0.967 \\
0.8 & 0.633 & 0.700 & 0.633 \\
0.9 & 0.067 & 0.067 & 0.067 \\
\bottomrule
\end{tabular}
\end{table}

All three models performed perfectly at thresholds of 0.5 and 0.6, with a slight dropoff at 0.7 and moderate degradation at 0.8. This validates our use of a 66th-percentile selection threshold and suggests that the system generally recommends questions for which students provide semantically valid answers—even under paraphrasing or stylistic variation.

\subsection*{Implications for Fairness and Instructional Validity}

These results highlight the importance of embedding model selection in open-ended educational systems. While overall performance remains robust across models, subtle differences in semantic sensitivity can affect how fairly and accurately student answers are interpreted. To mitigate bias, our implementation treats the canonical answer embedding as a flexible semantic anchor rather than a rigid template. This encourages conceptual alignment rather than surface-level matching and supports valid assessment across diverse linguistic styles.

In summary, KUL-Rec’s short-answer evaluation framework demonstrates resilience across embedding backbones, and its performance remains interpretable under a range of similarity thresholds. This supports its use in formative assessment settings that demand both semantic flexibility and pedagogical rigor.

\newpage
\section{Human Subjects Compliance and Survey Instrumentation}
\label{appendix:human_subjects}

This study was conducted under the approval of an Institutional Review Board (IRB) and determined to be exempt under 45 CFR 46.104(d)(1)(2), which covers educational research involving minimal risk. Participation in the adaptive quiz condition was voluntary, with students informed of the study prior to use. Signature requirements were waived, and all data were anonymized prior to analysis. No identifiable information was collected or retained.

\subsection*{Participant Instructions (as displayed in LMS)}

\textit{Welcome to Your Weekly Reading Assessment Quiz! This quiz is designed to assess your understanding of the key concepts, ideas, and themes from this week’s reading. Be sure to review the material carefully before starting, and answer each question to the best of your ability. Good luck!}

\subsection*{Post-Quiz Survey Prompts}

\begin{itemize}
    \item \textit{How much time did you spend on reviewing the readings this week before taking this quiz?}  
    (Less than 15 minutes, 15–30 minutes, 30–45 minutes, more than 45 minutes, or N/A)
    
    \item \textit{How would you rate the level of difficulty of this quiz?}  
    (Very difficult, somewhat difficult, neither easy nor difficult, somewhat easy, very easy)

    \item \textit{How helpful were these questions in consolidating the key points from the assigned readings?}  
    (Very helpful, somewhat helpful, neither helpful nor unhelpful, somewhat unhelpful, very unhelpful)
\end{itemize}

\subsection*{Interface and Delivery Platform}

Quizzes were distributed using the university's Canvas Learning Management System (LMS). A Python script using the Canvas API automated quiz assignment and submission. The interface mirrored standard Canvas quizzes and required no browser plugins or third-party tools.

\end{document}